\documentclass[10pt]{article}
\usepackage[dvips]{graphicx}
\begin{document}

\begin{center}
{\bf Nonlinear drift-diffusion model of gating in the fast Cl channel}
 \end{center}

  \begin{center} S. R. Vaccaro  \end{center}
  \begin{center} 
{\em Department of Physics, University of Adelaide, Adelaide, South Australia, 
5005, 
Australia} \\
  \end{center}
{\em E-mail address: svaccaro@physics.adelaide.edu.au}\\

{\bf Abstract}
   \begin{quotation}
The dynamics of the open or closed state region of an ion channel may be 
described by a probability density $p(x,t)$ which satisfies a Fokker-Planck 
equation. The closed state dwell-time distribution $f_c(t)$ derived from the 
Fokker-Planck equation with a nonlinear diffusion coefficient 
$D(x) \propto \exp(-\gamma x)$, $\gamma > 0$ and a linear ramp potential 
$U_c(x)$, is in good agreement with experimental data and it may be shown 
analytically that if $\gamma$ is sufficiently large, 
$f_c(t) \propto t^{-2 - \nu}$  for intermediate times,  where 
$\nu = U_c^{\prime}/\gamma  \approx -0.3$ for a fast Cl channel. 
The solution of a master equation which approximates  the Fokker-Planck 
equation exhibits an oscillation superimposed on the power law trend and can 
account  for an empirical rate-amplitude correlation that applies to 
several ion channels. 
\end{quotation}

	
\vspace{0.1in}

{\bf INTRODUCTION}

Ion channels are macromolecules which permit the conduction of ions across the 
membrane and are essential for metabolic cellular processes and information 
processing in the nervous system. The transition of a channel from the closed to 
the open state is regulated by the motion of one or more helical molecules which 
may depend on the membrane potential or the binding of a neurotransmitter to a 
receptor \cite{hh,hi}. The open and closed state dwell-time distributions 
obtained from the patch clamp recording of stochastic current pulses in ion 
channels may be represented by a finite sum of exponential functions of time 
\cite{ch}
 \begin{equation}
f(t) = \sum_{i = 1}^{N} a_i k_i \exp(-k_i t).
\label{fc}   \end{equation}
The discrete state Markov model assumes that the rate constants $k_i$ 
and the amplitudes $a_i$ may be derived from the transition rates
 between a small number $N$ of distinct conformational sub-states that form the 
open or closed state, and has been successful in describing gating current and
 dwell-time distributions in ion channels with the transition rates usually 
assumed to be independent. However, by assuming that the amplitude $a_i$ and the 
rate $k_i$ satisfy an empirical correlation $a_i \propto k_i^p$ where 
$p \approx 0.5$ \cite{mso1}, the resulting $f(t)$ exhibits an oscillation 
superimposed on a power law which provides an approximate fit to the 
dwell-time distribution obtained from some ion channels \cite{nn}.

By contrast to the discrete state Markov model, diffusion models 
assume that there are a large number of closed states, and are 
able to describe the approximate time course of gating currents \cite{sqb,le}, 
the intermediate power law behaviour of the dwell-time distribution 
$f_c(t) \propto t^{-1.5}$ when the diffusion coefficient is constant 
\cite{mso2,la,cj,gh1,gh2},  and $f_c(t) \propto t^{-2}$ when the approximately 
equal forward and backward  transition rates between neighbouring states 
decrease geometrically  away from the open state \cite{lfk,lieb}.  
An intermediate power law of the type 
$f_c(t) \propto t^{-2 + \alpha/2}$, where $\alpha$ is the index of anomalous 
diffusion ($\alpha = 1$ for normal diffusion) may be derived from a fractional 
diffusion model of ion channel gating and is in qualitative agreement with the 
data from a locust Ca-dependent BK channel when $\alpha = 0.14$ \cite{gh3}. 

The voltage dependence of the channel opening and  closing rate functions may be 
derived from the mean state residence time for an interacting diffusion regime 
\cite{gh1,gh2} or from an expression for the quasi-stationary diffusion current 
between the open and closed regions at each membrane surface, and in the latter 
case, the interaction between the open state probability and the membrane 
potential may be described by a Lagrangian (see Appendix).
It may be  shown that the closed state dwell-time distribution $f_c(t)$ 
derived from a Fokker-Planck equation with a nonlinear diffusion coefficient 
$D(x) = D_c \exp(-\gamma x)$, $\gamma > 0$ and a linear potential $U_c(x)$ is in 
good agreement with experimental data from a K and nACh channel and for 
intermediate times, $f_c(t) \propto t^{-1.5}$ when 
$\nu = U_c^{\prime}/\gamma  = -0.5$ where $U_c^{\prime} = {\partial 
U_c(x)}{\partial x}$ is a constant \cite{va}. In this paper, it is shown 
analytically that if $\gamma$ is sufficiently large, the solution of the 
Fokker-Planck equation has an intermediate power law approximation 
$f_c(t) \propto t^{-2 - \nu}$, and provides a good description of
 the data from a fast Cl channel when $\nu \approx -0.3$. 
The solution of the master equation approximation to the Fokker-Planck 
equation can also account for the empirical rate-amplitude correlation 
$a_i \propto k_i^p$ where $p \approx 0.65$ for a fast Cl channel.

\vspace{0.1in}

\begin{figure*}
\begin{center}
\includegraphics[width=0.8\textwidth]{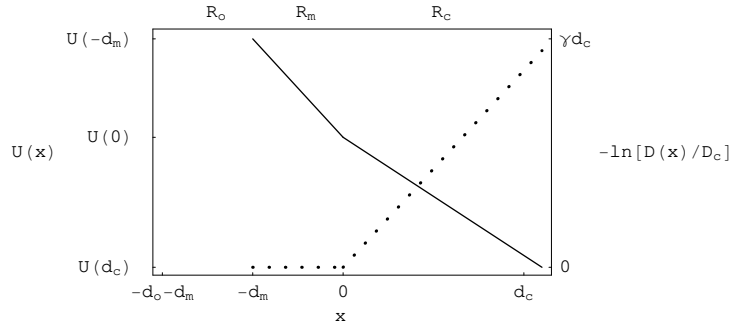}
\caption{
The potential function $U(x)$ (solid line) and $- \ln[D(x)/D_c]$ 
(dotted line) for the Brownian dynamics of a channel sensor with reaction 
coordinate $x$ in the closed state regions $R_m$ and $R_c$.
}
\end{center}
\end{figure*}

{\bf NONLINEAR DRIFT-DIFFUSION MODEL}

The opening of voltage and ligand gated channels is dependent on the 
configuration of a sensor which is comprised of one or more  helical 
molecules which may undergo rotation and translation between each surface of the 
membrane \cite{hi}. The states of the sensor are considered to form a linear 
chain and therefore, in a continuous model, each physical variable is a function 
of the one-dimensional reaction coordinate $x$. Positively charged residues on 
each sensor molecule are arranged in a regular array and interact with 
multiple  charged chemical groups on adjacent structures and the electrostatic
 environment to generate a sequence of energy wells and barriers 
\cite{llg}. It is assumed that the open state region 
$R_o$ ($-d_o - d_m \le x \le -d_m$) is adjacent to the closed state region for 
the sensor which is comprised of $R_m$ ($-d_m \le x \le 0$), 
where the diffusion coefficient $D(x)$ is a constant
 $D_m$, and  $R_c$ ($0 \le x \le d_c$) where the increase in barrier height 
between closed states in the direction away from the open state is represented
 by a nonlinear diffusion coefficient $D(x)$ (see Fig. 1). The continuous 
diffusion regime in $R_c$ may be approximated by discrete diffusion between a 
large number $N$ of states where the transition rates
$g_i = g_1 \sigma^{1 -  i}$, $b_i = b_1 \sigma^{1 -  i}$  for $i = 2$ to 
$N - 1$, $\sigma > 1$ (see Fig. 2), $D(x_i) \propto g_i$, 
$x_i = d_c(i - 1)/(N - 1)$ and therefore $D(x) \propto \exp(- \gamma x)$, 
$\gamma =  (N - 1) \ln \sigma/d_c > 0$. 
Qualitative agreement between gating current observed in K 
channels and that computed from a Fokker-Planck equation is obtained by
assuming that the diffusion coefficient is dependent on the reaction 
coordinate \cite{sqb}.

\begin{figure*}
\begin{center}
\includegraphics[width=0.8\textwidth]{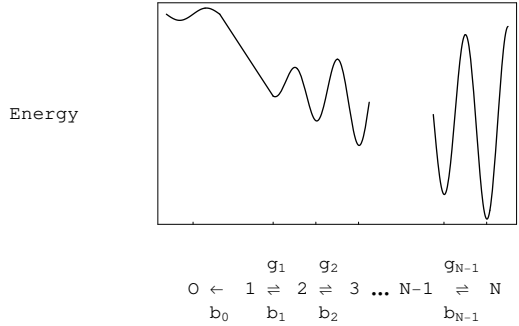}
\caption{
Energy level diagram for a Markov chain of N closed states of a channel sensor
with increasing barrier height and decreasing energy away from the open state O.
}
\end{center}
\end{figure*}

The probability density $p(x,t)$ of states of the sensor satisfies 
a Fokker-Planck equation \cite{kr,vk,ris} 
    \begin{equation}
\frac{\partial p(x,t)}{\partial t} = - \frac{\partial j(x,t)}{\partial x} =
\frac{\partial}{\partial x} \left(D(x) \left(\frac{\partial p(x,t)}{\partial x} 
+ 
\frac{\partial U(x)}{\partial x}  p(x,t) \right) \right),
\label{diff_gen} \end{equation}
where $j(x,t)$ is the probability current, and $U(x)$ is assumed to be a 
linear potential function, in units of $kT$ where $k$ is Boltzman's 
constant and $T$ is the absolute temperature. The Brownian motion of the sensor 
in $R_o$, $R_m$ and $R_c$ is a continuous generalization of thermally activated 
transitions between a finite number of closed states  and an open state in  
discrete diffusion models  of gating (see Figs. 1 and 2) \cite{mso2, cj, lieb}. 
An analytical solution of equation (\ref{diff_gen}) has been presented when 
the diffusion coefficient $D(x)$ and the potential function $U(x)$ are 
independent of $x$ \cite{gh1,gh2}. 

We shall assume that in the region $R_m$, $U_m(x)$ is dependent on the charge
$Q$ transferred across $R_m$ and the potential difference $V$ across the 
membrane relative to the external medium, the diffusion time
 $\tau_m = d_m^2/D_m$  is small relative to the mean time that the sensor 
resides in $R_c$, and that $p(-d_m,t) = 0$ at the boundary between 
the open and closed regions. Therefore, the uni-directional probability 
current is quasi-stationary and may be approximated by the expression  \cite{kr}
\begin{equation}
 j_m(t) =  - \frac{p(0,t) D_m}{\int_{-d_m}^0 \exp [U_m(x)- U_m(0)] dx}.
\label{jm} \end{equation}       

The transitions between closed states in $R_c$ are confined by the inner surface 
of the membrane, and therefore a reflecting boundary is imposed at $x = d_c$ 
\cite{mso1,cj,gh1,gh2}
 \begin{equation}
\frac{\partial p(x,t)}{\partial x} + U_c^{\prime} p(x,t) = 0,
\label{bc1} \end{equation}
where $U_c^{\prime}$ is a constant. The probability current at the interface 
between $R_m$ and  $R_c$ is continuous 
and thus
\begin{equation}
 j_c(0,t) = j_m(t)
\label{bc2} \end{equation}
where $j_m(t)$ is given by Eq. (\ref{jm}). 
It is assumed that, for each channel opening, the dwell time for the 
closed region begins when a sensor molecule is  transferred  across the region 
$R_m$ to the closed state at $x = 0$ in $R_c$, and thus $p(x,0) = \delta(x)$. 
 
        In the region $R_c$, Eq. (\ref{diff_gen}) may be expressed as 
\begin{equation}
\frac{\partial n(z,t)}{\partial t} = 
D_c \left(\frac{\partial^2 n(z,t)}{\partial z^2} + 
\frac{1}{z} \frac{\partial n(z,t)}{\partial z} -
 \frac{n(z,t)(\nu + 1)^2}{z^2} \right),
\label{diff_gen2} \end{equation}
where $z = z_0\exp(\gamma x/2)$, $z_0 = 2/\gamma$, 
$z_d = z_0 \exp(\gamma d_c/2)$, $\nu = U_c^{\prime}/\gamma$ and 
$n(z,t) = z^{\nu - 1} p(x,t)$. From the solution of 
Eq. (\ref{diff_gen2}) with the initial condition and the boundary 
conditions (\ref{bc1}) and (\ref{bc2}) using the method of 
Laplace transforms, it may be shown that the probability that the sensor is in 
the closed state region $R_c$ is 
\begin{equation}
 P_c(t) = \int_{0}^{d_c} p(x,t) dx = \sum_{i = 1}^{\infty} a_i \exp(-\omega_i 
t),
\label{sol1} \end{equation}
where $\omega_i = D_c \mu_i^2$, $\mu_i$ ($ < \mu_{i + 1}$) is a solution of the 
eigenvalue equation
 \begin{equation}
\frac{S_{\nu}(\mu_i,z_0)}{C_{\nu}(\mu_i,z_0)} = \frac{r_c}{\mu_i (z_d - z_0)},
\label{eigen_c} \end{equation}
$r_c = 2D_m [\exp(\gamma d_c/2) - 1]/\gamma D_c Y_m$, $Y_m = \int_{-d_m}^0 \exp 
[U_m(x) - U_m(0)] dx$, 
$C_{\nu}(\mu_i, z)$ and $S_{\nu}(\mu_i, z)$ are 
defined in terms of Bessel functions of the first kind
 \[C_{\nu}(\mu,z) = J_{-\nu}(\mu z_d) J_{\nu + 1}(\mu z) +
J_{\nu}(\mu z_d) J_{- \nu - 1}(\mu z),\]
\[S_{\nu}(\mu,z) = J_{-\nu}(\mu z_d) J_{\nu}(\mu z) - J_{\nu}(\mu z_d) J_{-
\nu}(\mu z),\] 
$a_i = 2/[1 + h_1(\mu_i) + h_2(\mu_i)]$ and
 \begin{equation}
h_1(\mu) = \frac{\mu (z_d - z_0)}{r_c C_{\nu}(\mu,z_0)} 
\frac{d[\mu S_{\nu}(\mu,z_0)]}{d \mu}, h_2(\mu) =
 \frac{1}{C_{\nu}(\mu,z_0)} \frac{d[\mu C_{\nu}(\mu,z_0)]}{d \mu}.
\label{g12} \end{equation}
 When $\nu$ is an integer, the solution of Eq. (\ref{diff_gen2}) may be 
expressed 
in terms of Bessel functions of the first and second kind. If $r_c$ is 
sufficiently small, it may be shown from Eq. (\ref{g12}) that $a_1 \approx 
1$, $a_i \approx 0$ for $i > 1$, the probability 
$P_c(t)  \approx \exp(- \omega_1 t)$ where 
$\omega_1 = D_m U_c^{\prime}/Y_m[1 - \exp(-U_c^{\prime} d_c)]$,
and therefore the solution accounts for  the exponential distribution of closed 
times described in slow K channels \cite{rsbv}.

The distribution function $f_c(t)= -dP_c(t)/dt$ obtained from 
 Eq. (\ref{sol1}) is also in good agreement with the approximate power law 
distributions of closed times from a K channel and a nACh channel when $\nu 
\approx -0.5$  \cite{va},  and a fast Cl channel when $\nu \approx -0.3$ (see 
Fig. 3). For $\gamma d_c \gg 1$, 
$r_c \gg 1$ and $-1 < \nu < 0$, the power 
law behaviour of $f_c(t) \propto t^{-2 - \nu}$, for intermediate times, may be 
derived by adopting a small argument approximation for $J_\nu(\mu z_0)$ 
and a large argument approximation for $J_\nu(\mu z_d)$ \cite{as},
\[
h_1(\mu) \approx \mu^2 (z_d - z_0)^2/r_c,
\]
\[
h_2(\mu) \approx \frac{r_c \Gamma(1+ \nu)(\mu/\gamma)^{-1-2 \nu}} {\theta 
\Gamma(-\nu)},
\]
\[ \theta \approx \frac{\cos[\mu z_d + \pi (\nu - 0.5)/2]}
         {\cos[\mu z_d - \pi (\nu + 1.5)/2]}.\]

\begin{figure*}
\begin{center}
\includegraphics[width=0.6\textwidth]{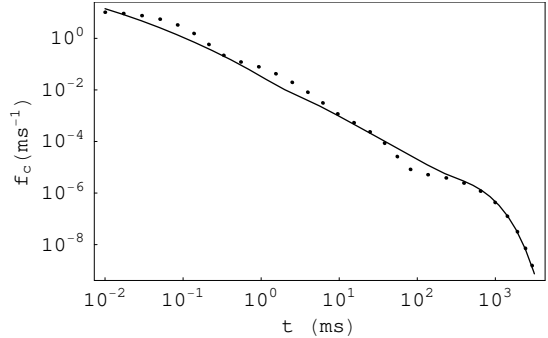}
\caption{
The closed state dwell-time distribution function $f_c(t)$ for a fast 
Cl channel \cite{bm} (dotted line) and the nonlinear
 drift-diffusion model (solid line) where $\tau_c = 1200$ ms, 
$r_c = 228$, $\gamma = 8/d_c$, $\nu = - 0.3$.}
\end{center}
\end{figure*}

There exists a positive integer $m$ such that for $i < m$, $h_2(\mu_i) \gg 
h_1(\mu_i)$
and hence for $t \gg 1/\omega_m$,
 \begin{equation}
P_c(t) \approx \frac{2 \pi A \tau_c^{\nu + 1/2}}{\Gamma(1+ \nu)} 
\sum_{i = 1}^{\infty} y_i^{1+ 2 \nu} \exp(-y_i^2 t), 
\label{pc1} \end{equation}
where $A = \theta \Gamma(-\nu) [2 \exp(\gamma d_c/2)]^{-1-2 \nu}/r_c \pi$, 
$\tau_c = (z_d - z_0)^2/D_c$,  and 
$y_i 
\approx \pi (i - 0.5)/\sqrt{\tau_c}$. For large $\tau_c$, 
$\Delta y_i = y_{i+1} - y_i = 2 y_1 \approx
\pi/\sqrt{\tau_c}$ is small, and if $t \ll \tau_c$, the sum of the infinite 
series may be approximated by the integral 
\[ 
\frac{\sqrt{\tau_c}}{\pi} \int_0^{\infty} y^{1 + 2 \nu} \exp{(- y^2 t)} dy  =
 \frac{\sqrt{\tau_c} \Gamma(1+ \nu)t^{-1- \nu}}{2 \pi} 
\]
 and thus
 \begin{equation}
P_c(t) \approx A(\frac{\tau_c}{t})^{1+ \nu}.
\label{pc2}  \end{equation}

When $\nu = -0.5$ this expression reduces to 
$P_c(t) \approx \sqrt{\tau_c/\pi t}/r_c$ which describes the power law 
approximation for a K and nACh ion channel \cite{va} whereas for a fast Cl
 channel, $\nu \approx -0.3$ and $P_c(t) \approx  0.1(\tau_c/t)^{0.7}/r_c \pi$.
If $t > \tau_c$, $P_c(t) \approx a_1 \exp(- D_c \mu_1^2 t)$ where
 $\mu_1$ is a solution of Eq. (\ref{eigen_c})  and 
therefore the continuous diffusion model describes the exponential tail
 that is often observed in the closed-time distribution. However,
 the patch clamp procedure has limited resolution whereas the solution of 
the Fokker-Planck equation includes an infinite number of high frequency 
components and therefore the agreement between the small time behaviour of the 
continuous model and the histogram data is only approximate. If $\gamma d_c \ll 
1$, adopting the large argument
 approximation for both $J_\nu(\mu z_0)$ and $J_{\nu}(\mu z_d)$, 
it may be shown that $P_c(t) \approx  \sqrt{\tau_c/\pi 
t}/r_c$ for intermediate  times, in agreement with the power law for the 
constant diffusion model ($\gamma = 0$) \cite{mso2,cj,gh1,gh2}. 

The mean closed time for the ion channel is  \cite{ch,gar}
\begin{equation}
T_c = \int_0^\infty t f_c(t) dt = \int_0^\infty P_c(t) dt,
\label{Tc}  \end{equation}
and from the solution (\ref{sol1}), when $U_m(x) = Q(V - V_f)(1 + x/d_m)/kT$ and 
$V_f$ is a constant,
\begin{equation}
\frac{1}{T_c} = \frac{D_m Q(V - V_f)}{d_m kT (1 - \exp[-Q(V - V_f)/kT])}
      \frac{U_c^{\prime}}{1 - \exp(-U_c^{\prime} d_c)},
\label{Tc2}  \end{equation}
and is independent of the mathematical form of the ion 
channel closed-time distribution. Eq. (\ref{Tc2}) may 
also be derived in the special case when a quasi-stationary state is attained in 
the closed region $R_c$ in a time $\ll T_c$ (see Appendix), and if 
$ U_c^{\prime} \rightarrow 0$, the expression reduces to that obtained
 from the constant diffusion model \cite{gh1}. The voltage dependence 
of the mean closed time determined from patch clamp data is 
generally in agreement with Eq. (\ref{Tc2}), but for some ion channels $T_c$ is 
only weakly dependent on $V$ \cite{rsbv,hza}.

Although the intermediate power law (\ref{pc2}) may be derived from the 
Fokker-Planck equation, the solution does not satisfy a 
rate-amplitude law $a_i \propto k_i^p$. Therefore, assuming that the ion 
channel sensor has a finite number of closed states \cite{hi}, we may consider a 
Markovian master equation which approximates Eq. (\ref{diff_gen}). 
If the channel sensor is able to undergo thermally 
activated transitions between N closed states and an open state (see Fig. 2), 
it may be assumed that the dynamics are described by a master equation 
\[ \frac{dp_1}{dt} = b_1 p_2 - (g_1 + b_0) p_1, \]
\begin{equation}
\frac{dp_i}{dt} = g_{i-1} p_{i-1} + b_{i+1} p_{i+1} - (g_i + b_{i-1}) p_i,(1 < i 
< N)
\label{pi}   \end{equation}
\[\frac{dp_N}{dt} = g_{N-1} p_{N-1} - b_{N-1} p_N,\]
where $p_i(t)$ is the probability of occupying the i-th closed state at time 
$t$, the transition rates are 
   \begin{equation} 
g_{i} = g_{i - 1}/\sigma_{i - 1}, b_{i} = b_{i - 1}/ \sigma_{i} 
\label{rates}   \end{equation}
for $1 < i < N$, $\sigma_1 \ldots \sigma_{N - 1}$  are the transition rate 
ratios, and $b_0$ is the rate between the first closed state and the open state.
The master equation model  with the transition rates (\ref{rates}) is a more 
general form of discrete diffusion models where $g_1 = b_1$, $\sigma_i = \sigma$ 
for each $i$ and either $\sigma = 1$  \cite{mso2} or $\sigma > 1$ \cite{lieb}.
As the difference in  reaction coordinate between states $\rightarrow 0$ and $N 
\rightarrow \infty$, the limit of the master equation when $\sigma_i = \sigma$ 
for each $i$, is a Fokker-Planck equation in the region $R_c$ \cite{gar,wei}. 

\begin{figure*}
\begin{center}
\includegraphics[width=0.7\textwidth]{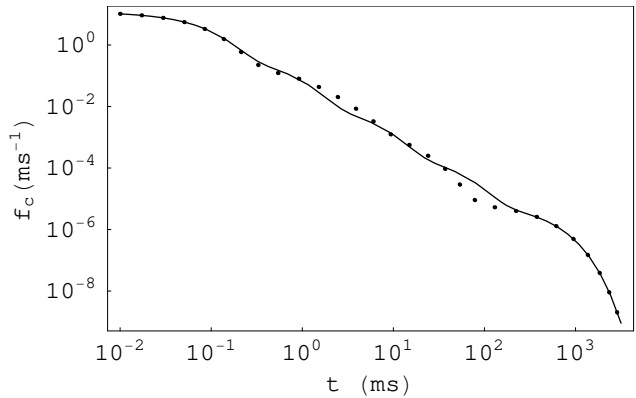}
\caption{
The closed-time distribution function $f_c(t)$ for a fast Cl channel 
 \cite{bm} (dotted line) and the master equation approximation to 
the nonlinear drift-diffusion  model (solid line), where $N = 5$, $b_0 = 3200$, 
$g_1 = 870.4$, $b_1 = 512$, $\sigma_i = 8$ for each $i$.
}
\end{center}
\end{figure*}

 The survival probability is given by
  \begin{equation} 
P_c(t) = \sum_{i = 1}^{N} p_i(t),
\label{prob}   \end{equation}
and the closed-time distribution function $f_c(t) = -dP_c/dt$ may be 
obtained by solving Eq. (\ref{pi}) with the initial condition 
$p_1(0) = 1$, $p_i(0) = 0$ for $i > 1$. The function $f_c(t)$ derived 
with uniform values for $\sigma_i$ provides a good 
fit to the data from a fast Cl channel (see Fig. 4), and 
 exhibits an oscillation superimposed on the power law 
trend for intermediate times. However, a better fit to 
the experimental data may be obtained by choosing non-uniform values for 
$\sigma_i$ (see Fig. 5). The values of $a_i$ and $k_i$ that are derived from 
the solution are comparable to those obtained experimentally and satisfy an 
approximate rate-amplitude correlation $a_i \propto k_i^p$ \cite{mso1} (see 
Fig. 6 where $p \approx 0.65$ for a fast Cl channel). The solution of 
the constant diffusion model ($\sigma_i = 1$) does not satisfy a 
rate-amplitude correlation but for sufficiently large $N$, $f_c(t) \propto t^{-
1.5}$ for intermediate times \cite{mso2}. 

\begin{figure*}
\begin{center}
\includegraphics[width=0.7\textwidth]{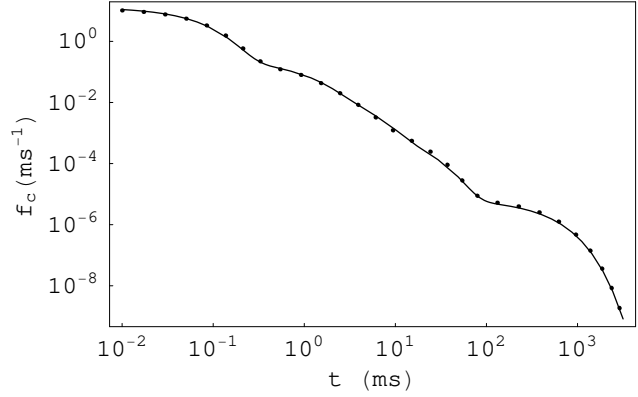}
\caption{
The closed-time distribution function $f_c(t)$ for a fast Cl channel 
\cite{bm} (dotted line) and the master equation approximation
 to the nonlinear drift-diffusion model (solid line), 
where $N = 5$, $b_0 = 3900$, $g_1 = 1200$, $b_1 = 380$, $\sigma_i = 
(15,3.5,4.7,21)$.
}
\end{center}
\end{figure*}

\begin{figure*}
\begin{center}
\includegraphics[width=0.7\textwidth]{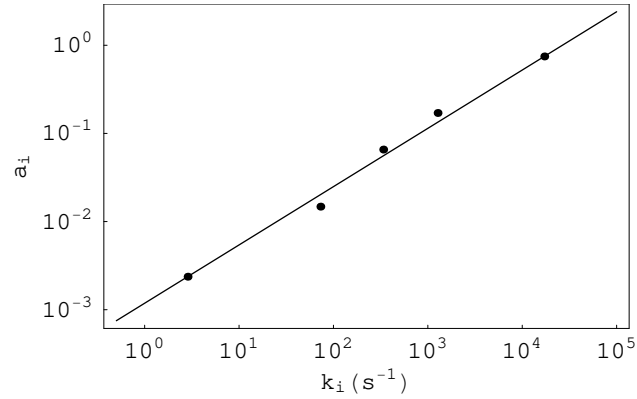}
\caption{
The amplitudes $a_i$ and the rates $k_i$ that are calculated from the master 
equation model of a fast Cl channel (see Fig. 5) satisfy the equation 
$a_i \propto k_i^p$ where $p \approx 0.65$ (solid line).
}
\end{center}
\end{figure*}

The rate-amplitude correlation and the intermediate power law that are observed 
for the closed-time distribution $f_c(t)$ of several types of ion channels 
\cite{mso1} may be derived by  considering Eq. (\ref{pi}) with $g_i = g_1 
\sigma^{1 -  i}$, $b_i = b_1 \sigma^{1 - i}$ for 
$i = 2$ to $N - 1$, $b_0 = b_1 \sigma$ where $\sigma > 1$ is sufficiently
 large and $p = \ln (b_i/g_{i+1})/\ln \sigma$. Using matrix methods, it
 may be shown that $P_c(t) = \sum_{i = 1}^{N} a_i  \exp(-k_i t)$ where 
$a_i/a_{i+1} \approx  b_i/g_{i+1} = \sigma^{p}$, $k_i/k_{i+1} \approx 
\sigma$, and hence $a_i \propto k_i^p$. For $t_i = 1/k_i$, 
\[ f_c(t_i) \approx \frac{t_i^{-p-1}}{e \sum_{j = 1}^{N} k_j} (1 + \sum_{j \neq 
i} 
T_j) \]
where $T_j = e(k_j/k_i)^{p+1}  \exp(- k_j/k_i)$, $\sum_{j \neq i} T_j \ll 1$
  and hence $f_c(t_i) \propto t_i^{-p-1}$ follows a general power law. 

{\bf DISCUSSION}

Discrete diffusion models of ion channel gating have dwell-time distributions 
which may be approximated by the intermediate  power law $t^{-3/2}$ 
when the transition rates  are constant \cite{mso2,cj}, and by 
$ t^{-2}$ when the forward and backward  transition 
rates between neighbouring states decrease geometrically  away from the 
open state \cite{lieb}. In this paper, we have considered a Fokker-Planck 
equation  which describes the dynamics of an ion channel sensor in the
 presence of a linear ramp potential $U_c(x)$ and an exponentially decreasing
 diffusion coefficient  $D(x) = D_c \exp(-\gamma x)$, and is a more 
general form of discrete and continuous diffusion models \cite{gh1,gh2}.
 The solution of the nonlinear diffusion model is dependent on the 
parameter $\nu = U_c^{\prime}/ \gamma$ and provides a good fit to 
the closed-time distribution function $f_c(t)$ for a delayed rectifier
 K channel and a nACh channel ($\nu \approx - 0.5$) and a fast Cl ion channel
($\nu \approx - 0.3$), and it may be shown analytically that for sufficiently 
large $\gamma$, $f_c(t) \propto t^{- 2 - \nu}$ for intermediate times.

Although the Fokker-Planck equation assumes a continuum of states, the ion 
channel sensor has a discrete structure and therefore the dynamics may be 
described by a Markovian master equation which approximates the 
nonlinear drift-diffusion equation. The distribution function $f_c(t)$ 
obtained from the solution to the master equation provides a good fit to the 
data from a fast Cl channel and exhibits an approximate rate-amplitude 
correlation $a_i \propto k_i^p$ where $p \approx 0.65$ \cite{mso1}.
 Therefore, a variation in the energy of closed states and an 
increase in the barrier height away from the open state 
are important factors in the closed-state dynamics of several ion channels.
 
\vspace{0.1in}
{\bf APPENDIX}

The channel opening and closing rate functions may be derived from an expression 
for the quasi-stationary diffusion current between the open and closed regions 
at each membrane surface when 
$p(-d_m,t) \propto P_o(t) = \int_{-d_m - d_0}^{-d_m} p(x,t) dx$ and 
$p(0,t) \propto P_c(t)$ (unpublished). If a quasi-stationary state is attained 
in the closed region $R_c$ in a time $\ll T_c$, and therefore corresponds to a 
small $r_c$ solution of the Fokker-Planck equation, from equations 
(\ref{sol1}) and (\ref{Tc}), we may write 
\begin{equation}
  p(0,t)   = 
\frac{P_c(t)}{ \int_{0}^{d_c} \exp[U_c(0) - U_c(x)] dx} 
\label{numc} \end{equation}
and $dP_c/dt = - P_c/T_c$ where 
\begin{equation}
\frac{1}{T_c} = \frac{D_m}{\int_{-d_m}^0 \exp[U_m(x)-U_m(0)] dx
\int_{0}^{d_c} \exp[U_c(0) - U_c(x)] dx }.
\label{alpha} \end{equation}
Similarly, if a quasi-stationary state is attained in the open region $R_o$ in a 
time $\ll T_o$, where $T_o$ is the mean open time, 
\begin{equation}
\frac{1}{T_o} = \frac{D_m}{\int_{-d_m}^0 \exp[U_m(x)- U_m(-d_m)] dx 
\int_{-d_m - d_0}^{-d_m} \exp[U_o(-d_m) - U_o(x)] dx }.
\label{beta} \end{equation}
Therefore,  each of the dwell-time distributions $f_c(t)$ and $f_o(t)$ is a 
single exponential function and  in agreement with the data from 
slow K channels \cite{rsbv}. 
If $U_m(x) = Q(V - V_f)(1 + x/d_m)/kT$ \cite{hh} and 
$U_c(x) = U_c(0) + U_c^{\prime} x$, the mean closed time $T_c$ reduces to Eq. 
(\ref{Tc2}) and a similar expression may be obtained for $T_o$.

The probability current between the open and closed state regions may be 
approximated by the expression \cite{kr}
\begin{equation}
 j_m(t) = - \frac{ D_m [p(0,t) \exp U_m(0) - p(-d_m,t) \exp U_m(-d_m)]}
{\int_{-d_m}^0 \exp U_m(x) dx}
\label{jm_gen} \end{equation} 
when the diffusion time $\tau_m \ll T_c$ or $T_o$. Therefore, assuming that 
$P_o(t) \approx 1 - P_c(t)$ and $P_f = \alpha/(\alpha + \beta)$ is the 
stationary value of $P_o$, where  $\alpha = 1/T_c$ is the mean opening rate,
 and $\beta = 1/T_c$ is the mean closing rate, from Eq. (\ref{numc}) and
 $p(-d_m,t) \propto P_o(t)$ we may write
\begin{equation}
\frac{dP_o(t)}{dt} = 
\alpha (1 - P_f) - \beta P_f  - (\alpha + \beta) (P_o(t) - P_f), 
\label{rateo} \end{equation}
a rate equation that describes the variation of K conductance in
 slow K channels ($g_K \propto P_o$)  \cite{rsbv}, and in delayed 
rectifier K channels assuming that the opening of the channel is
 determined by four identical and independent subunits  
 ($g_K \propto P_o^4$) \cite{hh}.

If $I_m$ is the macroscopic membrane K current across a membrane
 when each K channel is open, the  linear component of the
 ionic current is $I_m (P_o - P_f) = -C \dot{V}$. The nonlinear
 component of the K current and the other ionic currents through
 the membrane, such as Na,  are considered to be perturbations to the 
membrane potential. The net flow of ions across a membrane is dependent on the
K conductance which, in turn, is determined by the membrane potential. 
The voltage dependence of the rate functions (\ref{alpha}) and (\ref{beta}) 
may be expressed in terms of a Lagrangian L and dissipation function F which 
describe the interaction between the linear component of the ionic current and 
the quasi-stationary gating current between the closed and open region at each 
membrane surface 

 \[ L = \frac{\lambda (C \dot{V})^2}{2 I_m}  -
 \frac{C d_m kT}{Q} \int \frac{1 - \exp[U_m(-d_m)-U_m(0)]}{Y_m} dV, \]
\[F = \frac{\lambda (C \dot{V})^2}{2 I_m \tau}, \]
where $Y_c = \int_{0}^{d_c} \exp [U_c(0) - U_c(x)] dx$, 
$Y_o = \int_{- d_m - d_o}^{-d_m} \exp [U_o(-d_m) - U_o(x)] dx$,
$\lambda = (Y_c + Y_o)d_m kT/D_m Q$, $\tau = 1/(\alpha + \beta)$, 
the Lagrangian $L$ satisfies the equation 
    \[
\frac{d}{dt}(\frac{\partial L}{\partial \dot{q}}) -  
\frac{\partial L}{\partial q} + \frac{\partial F}{\partial \dot{q}} = 0,
\]
 and the canonical  coordinates $q = C(V - V_f)$ and $p = - \lambda (P_o - 
P_f)$.

\newpage


 \end{document}